\pdfoutput=1
\documentclass[aps, reprint, twocolumn,longbibliography,nofootinbib,floatfix]{revtex4-1}
\usepackage[latin1]{inputenc}
\usepackage{graphicx}
\usepackage{amsmath,amssymb,amstext,graphicx,bm,booktabs,hyperref}
\usepackage{fancyhdr}
\usepackage{microtype}
\usepackage{booktabs}

\usepackage{comment}
\usepackage{siunitx}
\usepackage{physics}

\usepackage{xcolor}
\usepackage{multirow}
\usepackage{placeins}

\newcommand*{\affaddr}[1]{#1}
\newcommand*{\affmark}[1][*]{\textsuperscript{#1}}
\fancypagestyle{alim}{\fancyhf{}\fancyfoot[R]{$^\dagger$Both authors contributed equally and are displayed in decreasing order of seniority.}}

\begin{document}

\title{Deep Reinforcement Learning for Efficient Measurement of Quantum Devices}

\author{V. Nguyen\affmark[$\dagger$1], S.B. Orbell\affmark[$\dagger$1], D.T. Lennon\affmark[1], H. Moon\affmark[1], F. Vigneau\affmark[1], L.C. Camenzind\affmark[4], L. Yu\affmark[4], D.M. Zumb\"{u}hl\affmark[4], G.A.D. Briggs\affmark[1], M.A. Osborne\affmark[2], D. Sejdinovic\affmark[3], and N. Ares\affmark[1]\\
\affaddr{\affmark[1]Department of Materials, University of Oxford, Oxford OX1 3PH, United Kingdom}\\
\affaddr{\affmark[2]Department of Engineering, University of Oxford, Oxford OX2 6ED, United Kingdom}\\
\affaddr{\affmark[3]Department of Statistics, University of Oxford, Oxford OX1 3LB, United Kingdom}\\
\affaddr{\affmark[4]Department of Physics, University of Basel, 4056 Basel, Switzerland}}

\begin{abstract}
Deep reinforcement learning is an emerging machine learning approach which can teach a computer to learn from their actions and rewards similar to the way humans learn from experience. It offers many advantages in automating decision processes to navigate large parameter spaces. This paper proposes a novel approach to the efficient measurement of quantum devices based on deep reinforcement learning. We focus on double quantum dot devices, demonstrating the fully automatic identification of specific transport features called bias triangles. Measurements targeting these features are difficult to automate, since bias triangles are found in otherwise featureless regions of the parameter space. Our algorithm identifies bias triangles in a mean time of less than $30$ minutes, and sometimes as little as $1$ minute. This approach, based on dueling deep Q-networks, can be adapted to a broad range of devices and target transport features. This is a crucial demonstration of the utility of deep reinforcement learning for decision making in the measurement and operation of quantum devices.

\end{abstract}

\maketitle

\date{\today{}}
\maketitle
\thispagestyle{alim}

\section*{Introduction}

Reinforcement learning (RL) is a neurobiologically-inspired machine learning paradigm where an RL agent will learn policies to successfully navigate or influence the environment. Neural network-based deep reinforcement learning (DRL) algorithms have proven to be very successful by surpassing human experts in domains such as the popular Atari 2600 games \cite{Mnih2015}, chess \cite{Silver2018}, and Go \cite{Silver2016}. RL algorithms are expected to advance the control of quantum devices \cite{August2018,Fosel2018,Bukov2018,Niu2019,Xu2019,Daraeizadeh2020,Herbert2018,Palittapongarnpim2017,An2019,Porotti2019a,Schuff2020,Wang2019,Wei2019,Gao2017,Barr2020,Deng2018,Carleo2017,Sordal2019}, because the models can be robust against noise and stochastic elements present in many physical systems and they can be trained without labelled data. However, the potential of deep reinforcement learning for the efficient measurement of quantum devices is still unexplored.

Semiconductor quantum dot devices are a promising candidate technology for the development of scalable quantum computing architectures. Singlet-triplet qubits encoded in double quantum dots \cite{Loss1997} have demonstrably long coherence times \cite{Malinowski2017}, as well as high one- and two-qubit gate fidelities \cite{Cerfontaine2019}. But quantum dot devices are subject to variability, and many measurements are required to characterise each device and find the conditions for qubit operation. Machine learning has been used to automate the tuning of devices from scratch, known as super coarse tuning \cite{Baart2016,Darulova2019a,Moon2020}, the identification of single or double quantum dot regimes, known as coarse tuning \cite{Zwolak, Zwolak2019}, and the tuning of the inter-dot tunnel couplings and other device parameters, referred to as fine tuning \cite{VanEsbroeck2020, Teske2019, Durrer2019}.

The efficient measurement and characterisation of quantum devices has been less explored so far. We have previously developed an efficient measurement algorithm for quantum dot devices combining a deep-generative model and an information-theoretic approach \cite{Lennon2019}. Other approaches have developed classification tools which are used in conjunction with numerical optimisation routines to navigate quantum dot current maps \cite{Zwolak, Kalantre2019, Teske2019}. These methods, however, fail when there are large areas in parameter space that do not exhibit transport features. To perform efficient measurements in these areas, which are often good for qubit operation, requires prior knowledge of the measurement landscape and a procedure to avoid over-fitting, i.e. a regularisation method.

In this paper, we propose to use DRL for the efficient measurement of a double quantum dot device. Our algorithm is capable of finding specific transport features, in particular bias triangles, surrounded by featureless areas in a current map. The state-of-the-art DRL decision agent is embedded within an efficient algorithmic workflow, resulting in significant reduction of the measurement time in comparison to existing methods. A convolutional neural network (CNN), a popular image classification tool~\cite{Krizhevsky2012,Lecun2015}, is used to identify the bias triangles.
This optimal decision process allows for the identification of promising areas of the parameter space without the need for human intervention. Fully automated approaches, such as the measurement algorithm presented here, could help to realise the full potential of spin qubits by addressing key difficulties in their scalability.


\begin{figure}[ht]
\begin{centering}
\includegraphics[width=0.99\columnwidth]{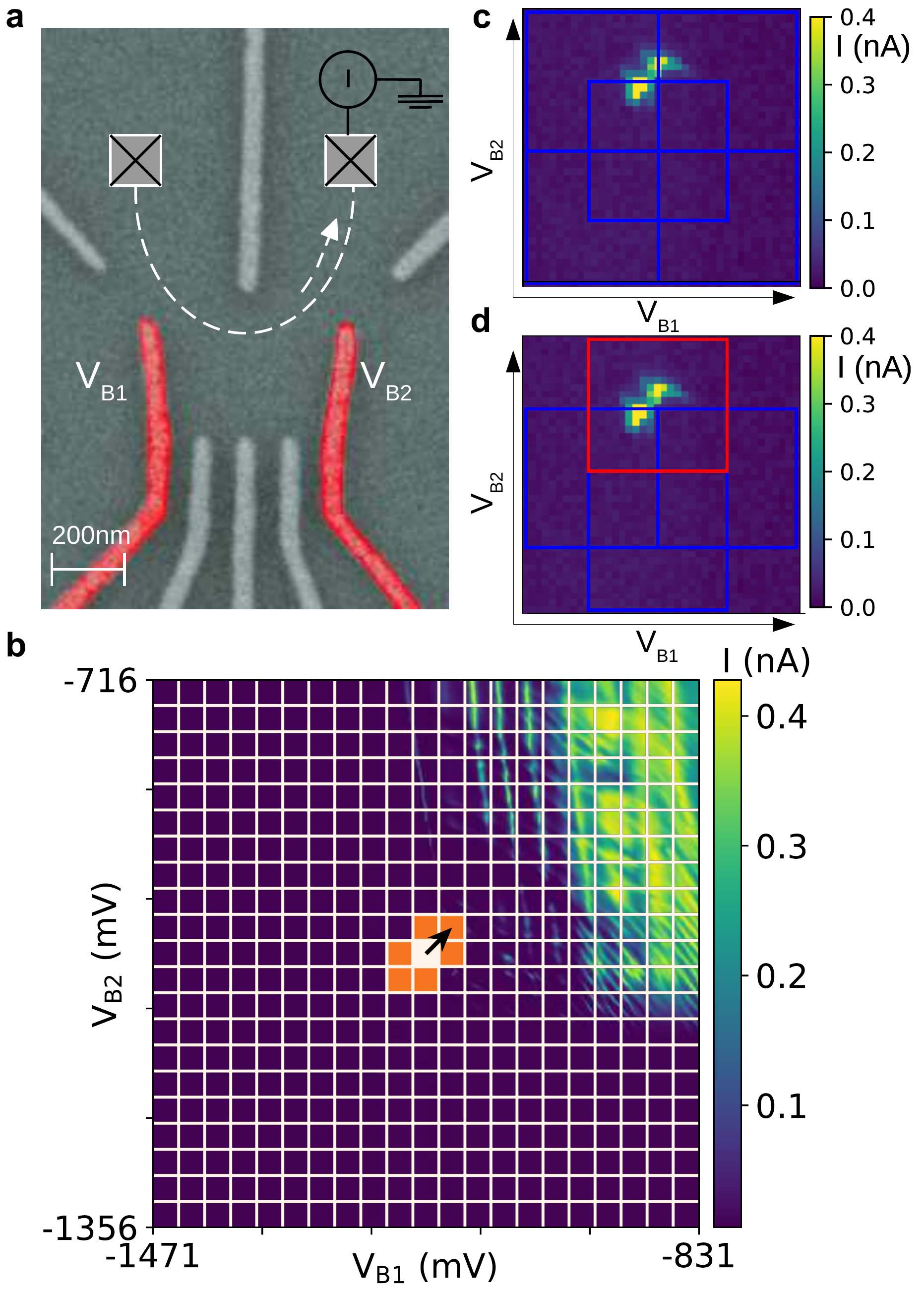}
\par\end{centering}
\centering{}\caption{ \textbf{Overview of device architecture and quantum dot environment}. \textbf{(a)} False-colour SEM image of a GaAs double quantum dot device. Barrier gates, labelled V$_{\mathrm{B1}}$ and V$_{\mathrm{B2}}$, are highlighted in red. The arrow represents the flow of current through the device between source and drain contacts. \textbf{(b)} A current map. The white grid represents the blocks available for investigation by the DRL agent. The DRL agent is initiated in a random block (state) indicated by a filled white square.
The filled orange blocks show the available action space for the DRL agent and the arrow shows a possible policy decision. \textbf{(c)} and \textbf{(d)} The nine sub-blocks defined within each block, a $\SI{32}{\milli\volt}$ $\times$ $\SI{32}{\milli\volt}$ window in gate voltage space, to calculate a statistical state vector. These sub-blocks are equal in gate voltage size, five of them are shown in (\textbf{c}) and four in (\textbf{d}). The red sub-block in (\textbf{d}) contains bias triangles.}
\label{fig:fig1}
\end{figure}

We focus on quantum dot devices that are electrostatically defined by Ti/Au gate electrodes fabricated on a GaAs/AlGaAs heterostructure (Fig. \ref{fig:fig1}\textbf{a}) \cite{Camenzind2018,Camenzind2019}. All the experiments were performed using GaAs double quantum dot devices at dilution refrigerator temperatures of approximately $\SI{30}{\milli\kelvin}$. The two-dimensional electron gas created at the interface of the two semiconductor materials is depleted by applying negative voltages to the gate electrodes. The confinement potential defines a double quantum dot which is controlled by these gate voltages and coupled to the electron reservoirs (the source and drain contacts). 
Depending on the combination of gate voltages, the double quantum dot can be in the `open', the `pinch-off' or the `single-electron transport' regime. In the `open' regime, an unimpeded current flows through the device. Conversely, when the current is completely blocked, the device is said to be in the `pinch-off' regime. In the `single-electron transport' regime, the current is maximal when the electrochemical potentials of each quantum dot are within the bias voltage window $\mathrm{V_{bias}}$ between source and drain contacts. 

Our algorithm interacts with a quantum dot environment within which our DRL decision agent operates to efficiently find the target transport features. The environment consists of states, defined by sets of measurements in gate voltage space, and a set of actions and rewards to navigate that space. This quantum dot environment has been developed based upon the OpenAI Gym interface \cite{Brockman2016a} (see Supplementary Information \ref{sup:sup_qde} for further details of the quantum dot environment's state, action and reward). Manual identification and characterisation of transport features requires a high-resolution measurement of a current map defined by, for example, barrier gate voltages \(\mathrm{V_{B1}}\) and \(\mathrm{V_{B2}}\) while keeping other gate voltages fixed, an example of which is shown in Fig.~\ref{fig:fig1}\textbf{b}. A super coarse tuning algorithm allows us to choose a set of promising gate voltages and focus on exploring the current map as a function of two gates, for example the two barrier gates~\cite{Moon2020}. This is the gate voltage space our DRL agent will navigate.


Our DRL algorithm takes the gate voltage coordinates found by our previous super coarse tuning algorithm \cite{Moon2020}, and divides the gate voltage space corresponding to the unmeasured current map into blocks. The size of the blocks is chosen such that they can fully contain bias triangles (blocks are shown as a white grid in Fig. \ref{fig:fig1}\textbf{b}). Devices with similar gate architectures often show bias triangles of similar sizes for a given $\mathrm{V_{bias}}$. The DRL agent is initiated in a random block. The agent acquires a reduced number of current measurements from this block and makes a decision on whether a high resolution measurement is required and on which block to explore next if bias triangles are not observed. The agent has a possible action space represented by a vector of length six; this means the agent can decide to acquire measurements in any of the four contiguous blocks (`up', `down', `left' or `right') or in the two diagonal blocks that permit the agent to efficiently move between the `open' and the `pinch-off' transport regimes. These blocks correspond to an increase or decrease of both gate voltages, which strongly modulates the current through the device. The remaining two diagonal blocks, which correspond to a decrease of one gate voltage and an increase of the other, do not often lead to such significant changes in the transport regime and are thus not included in the agent's action space to maximise the efficiency of the algorithm. The DRL agent can be efficiently trained using current maps already recorded from many other devices. This is because their transport features are sufficiently similar, even though the gate voltage values at which they are observed vary for different devices. 

The decision of which block to explore next is based on the current measurements acquired by the DRL agent in a given block. The block is divided into nine sub-blocks (Fig. \ref{fig:fig1} \textbf{c} and \textbf{d}) and the mean \(\mu\) and standard deviation \(\sigma\) of the current measurements corresponding to each sub-block are calculated. These statistical values, constituting an $18$-element vector, provide the agent with information of its probable location in the current map. The statistical state vector or state representation vector enables the DRL decision agent to abstract knowledge about the transport regime, distinguishing between `open', `pinch-off', and `single-electron transport' regimes, with a reduced number of measurements. In this way, the state vector defines a state in the quantum dot environment. 

This statistical approach, compared to the alternative of using CNNs to evaluate acquired measurements, makes the agent less prone to over-fitting during training and more robust to experimental noise. To decide whether the agent has found bias triangles in a given block, the algorithm uses a CNN as a binary classification tool. Combining a state representation based on measurement statistics and CNNs in a reinforcement learning framework which makes use of the experience of the agent navigating similar environments during training, our algorithm provides a decision process for efficient measurement without human intervention.

\section*{Results}

\begin{figure*}[th!]
\begin{centering} 
\includegraphics[width=1.95\columnwidth]{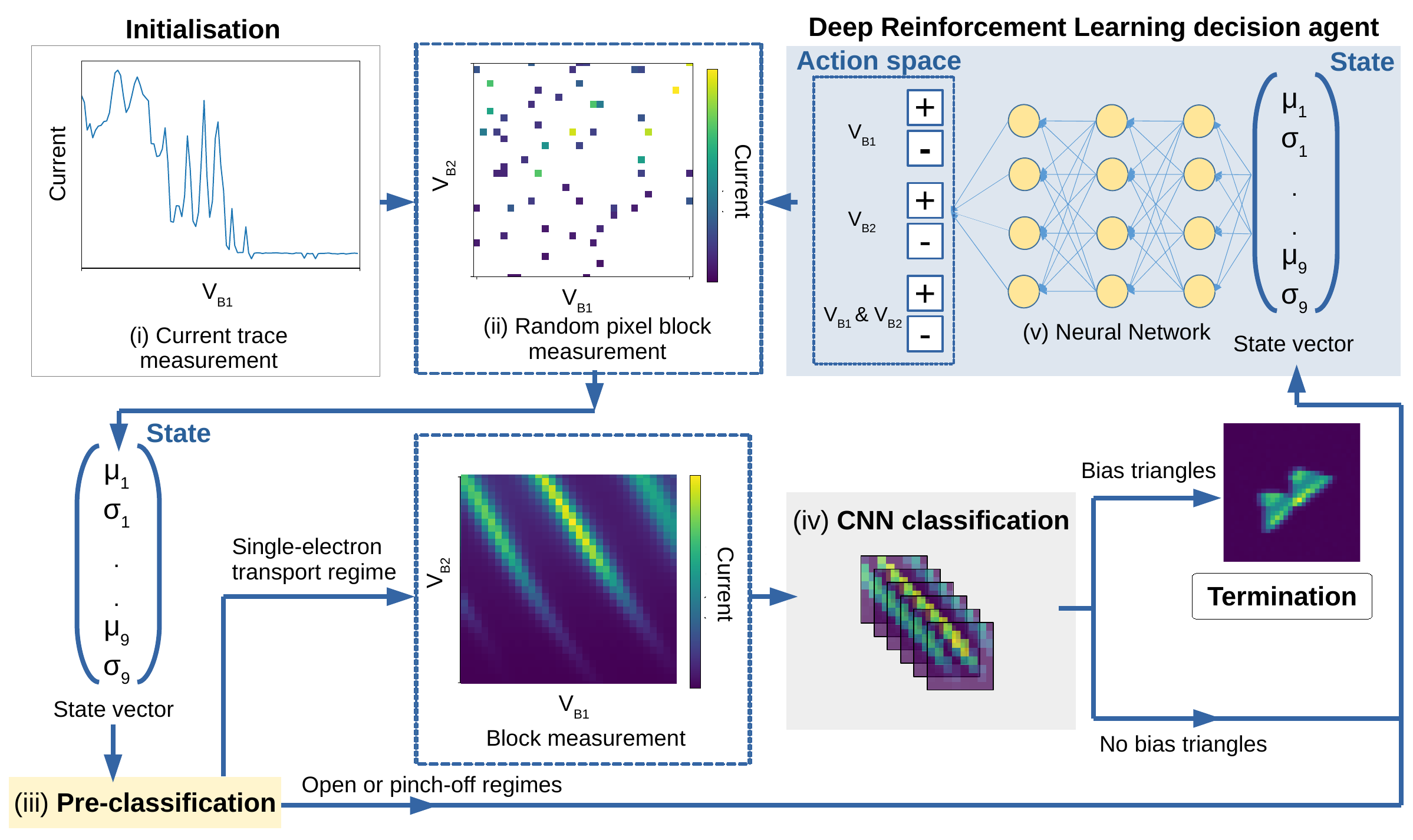}\\
\par\end{centering}
\centering{}\caption{\textbf{Schematic depicting the algorithmic workflow.} (See main text for a full description) In the initialisation stage, starting from the gate voltages coordinates proposed by a coarse tuning algorithm, the algorithm measures low-resolution current traces as a function of \(\mathrm{V_{B1}}\) (\(\mathrm{V_{B2}}\)) with \(\mathrm{V_{B2}}\) (\(\mathrm{V_{B1}}\)) set to the maximum voltage given by the gate voltage window of interest (\textbf{i}). The algorithm then performs a random pixel measurement in the block corresponding to the proposed starting gate voltages (\textbf{ii}). In this measurement, mean current values and standard deviation are calculated for 9 sub-blocks within the block until convergence. The statistical state representation vector (state vector) obtained is then assessed by the pre-classification stage (\textbf{iii}). If the mean current value corresponding to any of the sub-blocks falls within threshold values given by the initialisation stage, then the block is pre-classified as corresponding to a possible single-electron transport regime. In this case, the block is explored further by performing a high-resolution scan. This block measurement is normalised and input into a CNN binary classification algorithm (\textbf{iv}). If the CNN identifies bias triangles, then the algorithm terminates. If either the pre-classifier or the CNN-classifier rejects a block, then the state vector is input into the DRL decision agent  (\textbf{v}). The decision agent subsequently selects an action on the gate voltages which determines the next block to measure via the random pixel method. \label{fig:AlgorithmFlowchart}}
\end{figure*}

\subsection*{Description of the algorithm}

The algorithm is comprised of different modules for classification and decision making (Fig. \ref{fig:AlgorithmFlowchart}). In the initialisation stage, two low-resolution current traces are acquired by the algorithm as a function of \(\mathrm{V_{B1}}\) (\(\mathrm{V_{B2}}\)) with \(\mathrm{V_{B2}}\) (\(\mathrm{V_{B1}}\)) set to the maximum voltage given by the gate voltage window to be explored. The algorithm extracts from these measurements the maximum and minimum current values and its standard deviation, which will be used in a later stage by the classification modules. The gate voltage regions we explore are delimited by a $\SI{640}{\milli\volt}$ $\times$ $\SI{640}{\milli\volt}$ window centred in the gate voltage coordinates proposed by a super coarse tuning algorithm, as mentioned in the Introduction, and the current traces in this stage have a resolution of $\SI{6.4}{\milli\volt}$. 

The gate voltage region is divided in $\SI{32}{\milli\volt}$ $\times$ $\SI{32}{\milli\volt}$ blocks and the agent is initialised in a randomly selected block. The algorithm takes random pixel measurements of current within this block. Each pixel is $\SI{1}{\milli\volt}$ $\times$ $\SI{1}{\milli\volt}$. As these measurements are performed, the algorithm estimates the 18-dimensional state vector given by \(\mu\) and \(\sigma\) for each of the $9$ sub-blocks in which the block is divided. Pixels are sampled randomly from the block until the statistics from the state representation have converged. Convergence is generally achieved after sampling fewer than \(100\) pixels, significantly less than the \(1024\) pixels in a block (see Supplementary Information \ref{sup:sup_random_pixel} for the convergence curves and the convergence criterion).

\begin{figure}[ht]
\begin{centering}
\includegraphics[width=\columnwidth]{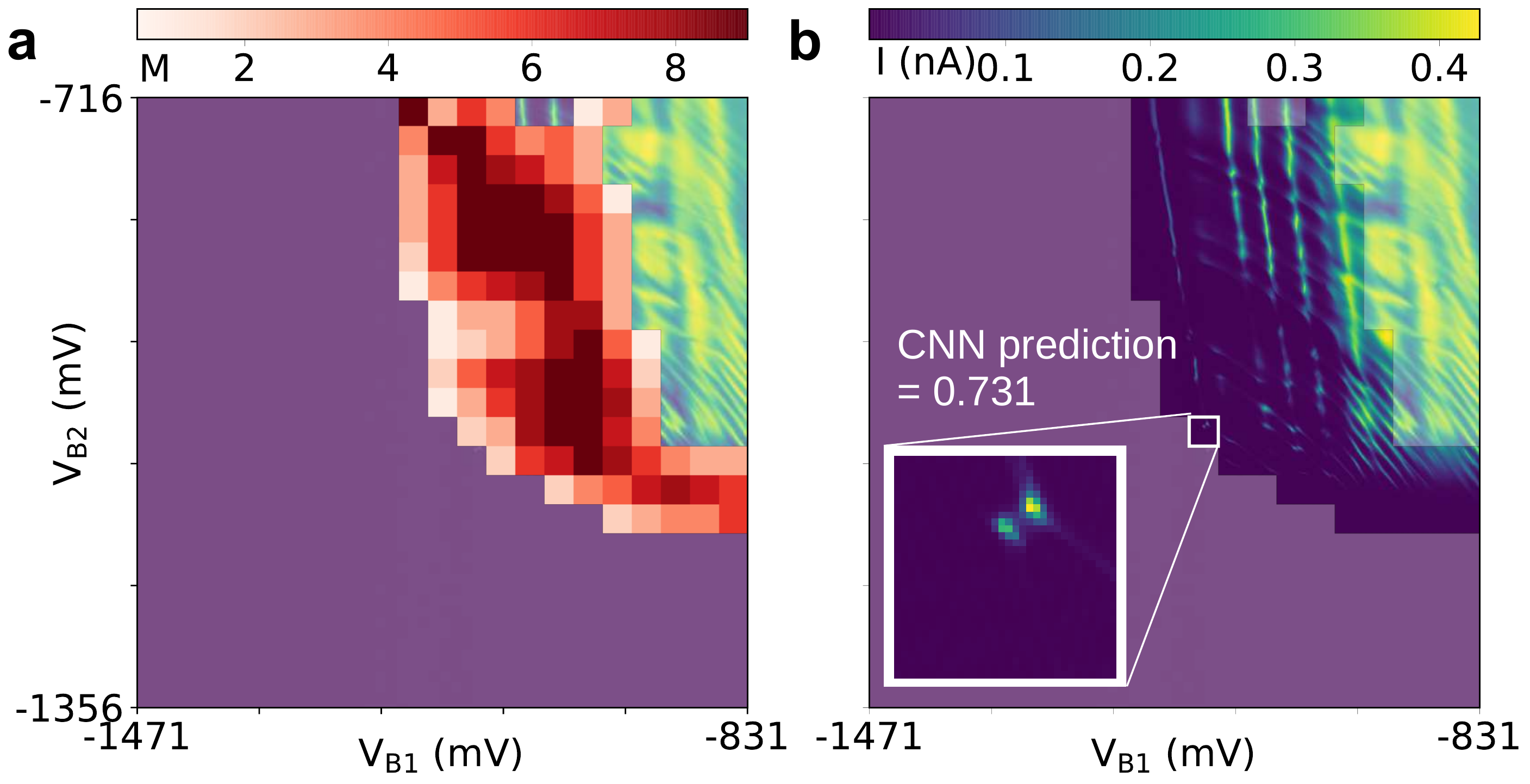}
\par\end{centering}
\centering{}\caption{\textbf{Classification tools.} \textbf{(a)} Example of blocks considered by the pre-classifier as corresponding to the `single-electron transport' regime overlaid on the corresponding current map. The colour-bar represents the number (M), out of nine, of sub-blocks which were not rejected by the pre-classification stage. \textbf{(b)} Blocks in \textbf{(a)}, displaying features corresponding to the `single-electron transport' regime, overlaid on the corresponding current map. Inset: A block displaying bias triangles and the corresponding output value of the CNN binary classifier. 
\label{fig:classifiers}}
\end{figure}

The state vector is first evaluated by a pre-classification module. A block is considered to correspond to the single-electron transport regime if any of the \(\mu\) values is within \(0.01\) to \(0.3\) times the maximum current range detected in the initialisation stage (see Supplementary Information \ref{sup:sup_random_pixel} for further details about the design of the pre-classifier). We have found that the choice of current range does not have a significant impact in the performance of the algorithm. If the pre-classifier identifies the block as corresponding to the single-electron transport regime, a high resolution current measurement (\(1024\) pixels, $\SI{1}{\milli\volt}$ $\times$ $\SI{1}{\milli\volt}$ resolution) of the block is acquired. 
This block measurement is normalised and evaluated by a CNN binary classifier. For any output value greater than \(0.5\), the block is identified as containing bias triangles. If bias triangles are identified within the block, the algorithm is terminated. Fig. \ref{fig:classifiers}\textbf{a} shows the blocks in a current map that would be identified by the pre-classifier as corresponding to the single-electron transport regime, while Fig. \ref{fig:classifiers}\textbf{b} shows the blocks that would be evaluated by the CNN binary classification to determine if bias triangles are observed (See Supplementary Information \ref{sup:sup_cnn} for a summary of the CNN's architecture and its training).

If the pre-classifier considers the block to correspond to the `open' or `pinch-off' regimes, or if the CNN does not identify bias triangles within the block, the DRL agent has to decide which block to explore next. With this objective, the state vector is normalised using the variance and mean current values obtained in the initialisation stage, and fed into a deep neural network which controls the DRL decision agent. The agent will then propose an action which it expects will lead to the highest long-term reward. This action $a_{t}$, given by \(a_{t}=\arg\max_{a'}Q^{\pi}(s_{t},a')\), is the action which maximises the $Q$-function for the agent's stochastic policy $\pi$ in the state-action pair $(s_{t},a')$ at time $t$. The $Q$-function measures the value of choosing an action $a'$ when in state $s_{t}$ and therefore the action $a_{t}$ represents the agent's prediction for the most efficient route to bias triangles. In our quantum dot environment setting, the action determines the next block to explore and the algorithm begins a new iteration.

\subsection*{The deep reinforcement learning agent}

Our algorithm makes use of the deep Q-learning framework which uses deep neural networks to approximate the $Q$-function \cite{Mnih2015}. The $Q$-function is defined by \(Q^{\pi}(s_{t},a_{t}) = \mathbb{E}\left[R_{t}|s = s_{t}, a = a_{t}, \pi \right]\), which gives an expected reward $R_{t}$ for a chosen action $a_{t}$ taken by an agent with a policy $\pi$ in the state $s_{t}$. This expected reward is defined as $R_t=\sum_{\tau=t}^\infty  \gamma^{\tau-t} r_\tau$, where $\gamma \in [0, 1]$ is a discount factor that trades-off the importance of immediate rewards $r_{t}$, and future rewards $r_{\tau>t}$. The agent aims to maximize $R_t$ via the $Q^{\pi}(s_{t},a_{t})$ learnt by the neural network. 
In particular, we chose to implement the dueling deep Q-network (dueling DQN) \cite{Wang2016} architecture for our DRL decision agent. This architecture factors the neural network into two entirely separate estimators for the state-value function and the state-dependent action advantage function \cite{Wang2016}. The state-value function, \(V^{\pi}(s_{t}) = \mathbb{E}_{a_{t} \sim \pi(s_{t})}\left[Q^{\pi}(s_{t},a_{t})\right]\) gives a measure for how valuable it is, for an agent with a stochastic policy $\pi$ in the search for a promising reward, to be in a given state $s_{t}$. The state-dependent action advantage function \cite{Wang2016} gives a relative measure of the importance of each action, given by \(A^{\pi}(s_{t},a_{t}) = Q^{\pi}(s_{t},a_{t}) - V^{\pi}(s_{t})\). In dueling DQN, when combining the state-value function and the state-dependent action advantage function, it is crucial to ensure that given $Q$ we can recover $V^{\pi}(s_{t})$ and $A^{\pi}(s_{t},a_{t})$ uniquely. For this purpose, the advantage function estimator is forced to be zero at the chosen action $a_{t}$ \cite{Wang2016}. 
This approach allows the agent, through the estimation of $V^{\pi}(s_{t})$, to learn the value of certain states in terms of their potential to guide the agent to a promising reward. 
This is particularly beneficial in our case, since different state vectors can correspond to the same transport regime and thus be equally valuable in the search of bias triangles. Consequently, the most beneficial action in these states would often coincide. For example, in most states corresponding to the `pinch-off' regime, the most beneficial action is often to increase both gate voltages.

To train the DRL agent, we designed a reward function to ensure that the agent would learn to efficiently locate bias triangles. To this end, during training, the agent is rewarded for the detection of bias triangles and penalised for the number of blocks explored or measured in a single algorithm run, $N$. The reward $r=+10$ is assigned to the blocks exhibiting bias triangles. Other blocks are assigned $r=-1$. During training, the maximum number of blocks that could be measured in a given run, $N_\mathrm{max}$, is set to $300$. If after $N_{\mathrm{max}}$ block measurements the agent had not found bias triangles, the algorithm is terminated and the agent is punished with $r=-10$  (see Supplementary Information \ref{sup:sup_qde} for further details regarding the design of the reward function). In other words, $N_\mathrm{{max}}$ determines how far from the starting block the agent can reach in gate voltage space, as it can only explore contiguous blocks. 

We trained the dueling DQN (DRL decision agent) using the prioritised experience replay method \cite{Schaul2016} from a memory buffer. This method ensures that successful policy decisions are replayed more frequently in the DRL agent's learning process. The agent does not benefit from an ordered sequence of episodes during learning, yet it is able to learn from rare but highly successful policy decisions and it is less likely to settle in local minima of the decision policy. We trained the agent over $10000$ episodes (algorithm runs), each time initialised in a random block for $4$ different current maps which were previously recorded. The training takes less than an hour on a single CPU.

\subsection*{Experimental results}

\begin{figure*}[!ht]

\includegraphics[width=2\columnwidth]{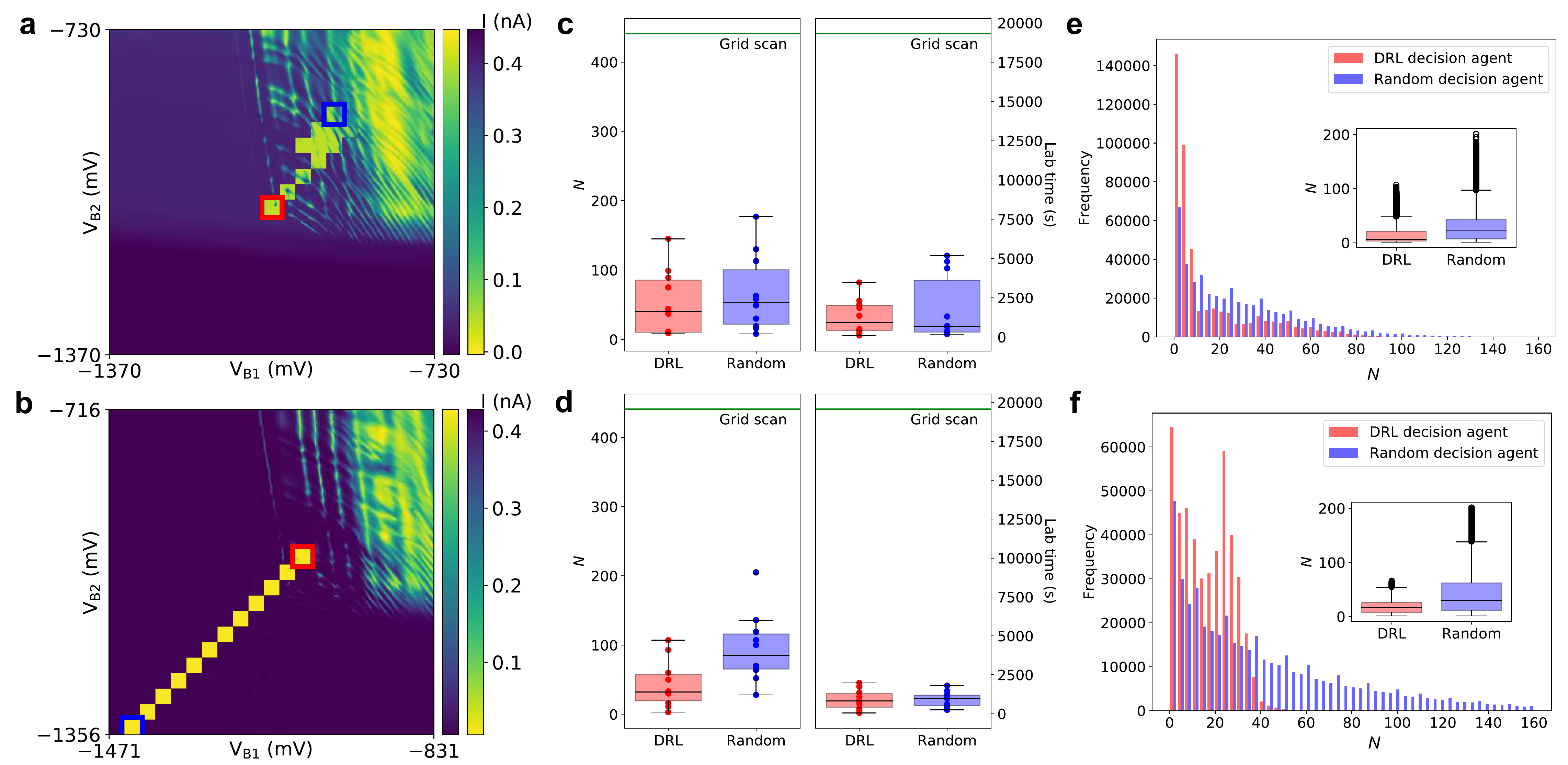}

\caption{\textbf{Performance benchmark.} 
\textbf{(a, b)} Example trajectories of the DRL agent in gate voltage space. \textbf{a} (\textbf{b}) Corresponds to region I (II).
The trajectories are indicated inverting the colour scale of the current map for the blocks measured by the algorithm. The current map measured by the grid scan method is displayed for illustrative purposes and it is not seen by the DRL agent. The blue and red squares indicate the start and end of the trajectory, respectively. \textbf{(c, d)} Real-time performance corresponding to the grid scan method (green line), the algorithm with a random decision agent (blue) and the algorithm with a DRL decision agent (red). The box plots indicate the laboratory time and the corresponding number of blocks explored, $N$, for regions of the gate voltage space I and II in \textbf{c} and \textbf{d}, respectively. The results of all 10 runs for both agents in each regime are plotted as points. The central line of the box plot corresponds to $\bar{N}$, while the upper and lower boundaries of the box display the upper ($Q3$) and lower ($Q1$) quartiles. The minimum and maximum whisker bars display ($Q1 - 1.5 \times IQR$) and ($Q3 + 1.5 \times IQR$) respectively, where $IQR$ is the interquartile range. \textbf{(e, f)} Histograms of values of $N$ for the random and DRL decision agents over $10$ algorithm runs for each region, I (\textbf{e}) and II (\textbf{f}). This performance test was performed offline. The insets show the box plots, indicating the quartiles and $\bar{N}$ values for the DRL and random agents. In the inset only the outlier points are plotted. \label{fig:PerformanceBenchmark}}
\end{figure*}

We demonstrate the real-time (`online') performance of our algorithm in a double quantum dot device. The algorithm performance is evaluated according to the number of blocks explored in an algorithm run, $N$, which is equal to the number of blocks explored to successfully identify bias triangles unless $N=N_\mathrm{max}$, and according to the laboratory time spent in this task. For training and testing the algorithm's performance we use different devices, both similar to the device shown in Fig. \ref{fig:fig1}\textbf{a}. We ran the DRL algorithm in two different regions of gate voltage space, I and II, which are centred in the coordinates from our super coarse tuning algorithm \cite{Moon2020}. We ran the algorithm $10$ times in each region. The DRL agent was initiated in a different block for every run, sampled uniformly at random. From these repeated runs, we can estimate the median $\bar{N}$ of the distribution of values of $N$ obtained for a given region. We can also estimate $(L,U)$, where $L$ and $U$ are the lower and upper deciles of the distribution. To identify bias triangles, the DRL agent required $\bar{N} = 40$ $(9, 104)$ for region I and $\bar{N} = 32$ $(10, 94)$ for region II. In both regions considered, our algorithm efficiently located bias triangles in a mean time of $30$ minutes and, on one occasion, in less than $1$ minute. This is an order of magnitude improvement in measurement efficiency compared to the laboratory time required to acquire a current map with the grid scan method, i.e. measuring the current while sweeping \(\mathrm{V_{B2}}\) and stepping \(\mathrm{V_{B1}}\), which is approximately $5.5$ hours with pixel resolution ($\SI{1}{\milli\volt}$ $\times$ $\SI{1}{\milli\volt}$ resolution). Using a single CPU of a standard desktop computer, the algorithm is not limited by computation time. It can thus be run with the computing resources available in most laboratories.

Example trajectories of the agent within the gate voltage space give an insight into the transport properties that the agent has implicitly learnt from its environment. When initiated in a transport regime corresponding to pinch off (low current), the agent reduces the magnitude of the negative voltage applied to the gate electrodes, as humans experts would do (Fig. \ref{fig:PerformanceBenchmark}\textbf{a}). Conversely, when initiated in a transport regime corresponding to higher currents, the agent increases the magnitude of the negative voltage applied to the gate electrodes (Fig. \ref{fig:PerformanceBenchmark}\textbf{b}). The policy thus leads to block measurements in the areas of gate voltage space where bias triangles are usually located.

We have performed an ablation study. Ablation studies are used to identify the relative contribution of different algorithm components. In this case, our aim is to determine the benefit of using a DRL agent. We thus produced an algorithm in which the DRL decision agent was replaced with a random decision agent. We compared its performance with the DRL algorithm. The random agent selects an action, sampled uniformly and randomly. The QDE's action space is six-dimensional except in instances where the agent is in a state (block) along the edges (five-dimensional action space) and in the corners (four-dimensional action space) of the gate voltage window considered. This measurement strategy is similar to a random walk within the gate voltage space, but unlike a pure random walk strategy, it will not measure the same block twice. The random decision agent's measurement run will be terminated when the CNN classifies a block measurement as containing bias triangles. The random agent was initialised in the same random positions as the DRL agent so that a fair comparison could be made between their performances. We performed $10$ runs of each algorithm in each of the two different regions of parameter space considered in this work, I and II (Fig. \ref{fig:PerformanceBenchmark} \textbf{c} and \textbf{d}). The DRL agent outperforms the random decision agent in the value of $\bar{N}$, and thus in the laboratory time required to successfully identify bias triangles. Note that the relation between $\bar{N}$ and the laboratory time is not linear, as high-resolution block measurements are only performed for each block classified as corresponding to the single-electron transport regime by the pre-classification stage.

In region II, the random agent requires $\bar{N}$ equal to $85$ $(50, 143)$ which is approximately 2.6 times larger than the $\bar{N}$ corresponding to the DRL agent (see Supplementary Information \ref{sup:sup_drl_agent} for the value of $\bar{N}$ in region I and corresponding lab times). The good performance of the random decision agent can be explained by its use of the pre-classifier, which makes the random search efficient. The random decision agent is an order of magnitude quicker than the grid scan method.  

To test the statistical significance of the DRL agent's advantage, we have tested the performance of both algorithms in a much larger number of runs. To perform this statistical convergence test would have been too costly in laboratory time, so we used previously recorded current maps, which were measured by the grid scan method. We will call this performance test `offline', as opposed to `online' in the case of real-time measurements. By initiating both agents $1024$ times in each of the blocks in I and II, we obtained a histogram of the $N$ blocks measured to successfully terminate the algorithm (see Fig. \ref{fig:PerformanceBenchmark}\textbf{e} and \textbf{f} for I and II, respectively). We observe a higher number of runs for which the DRL algorithm performed fewer block measurements for successful termination. In region II, the DRL agent requires $\bar{N}$ of $17$ $(2, 31)$, while the $\bar{N}$ for the random agent is $30$ $(3, 101)$ (see Supplementary Information \ref{sup:sup_drl_agent} for the value of $\bar{N}$ for region I). Our results suggest that the DRL advantage is statistically significant. The two-tailed Wilcoxon signed rank test \cite{Wilcoxon1945} allows us to make a statistical comparison of the two distributions corresponding to the DRL and the random agent. We have applied this test to the offline performance for regions I and II (see Supplementary Information \ref{sup:sup_drl_agent} for the results of the Wilcoxon signed rank test applied to the online performance test, for which critical values for the test threshold are used instead of assuming a normal approximation, given the number of algorithm runs is below 20). The two-tailed Wilcoxon signed rank test yields a p-value $< 0.001$ for both regions. This means that the null hypothesis, stating there is no difference in the median performance between the two agents, can be rejected. In addition, the median of the differences $(\bar{N}_{\mathrm{DRL}}-\bar{N}_{\mathrm{Random}})$, estimated using the one-tailed Wilcoxon signed rank test, is less than zero. We can therefore confirm that the DRL agent offers a statistically significant advantage over the random agent.

\begin{figure}
\textbf{Region I} \hspace{2.5cm} \textbf{Region II} \\
\begin{centering}
\includegraphics[width=\columnwidth]{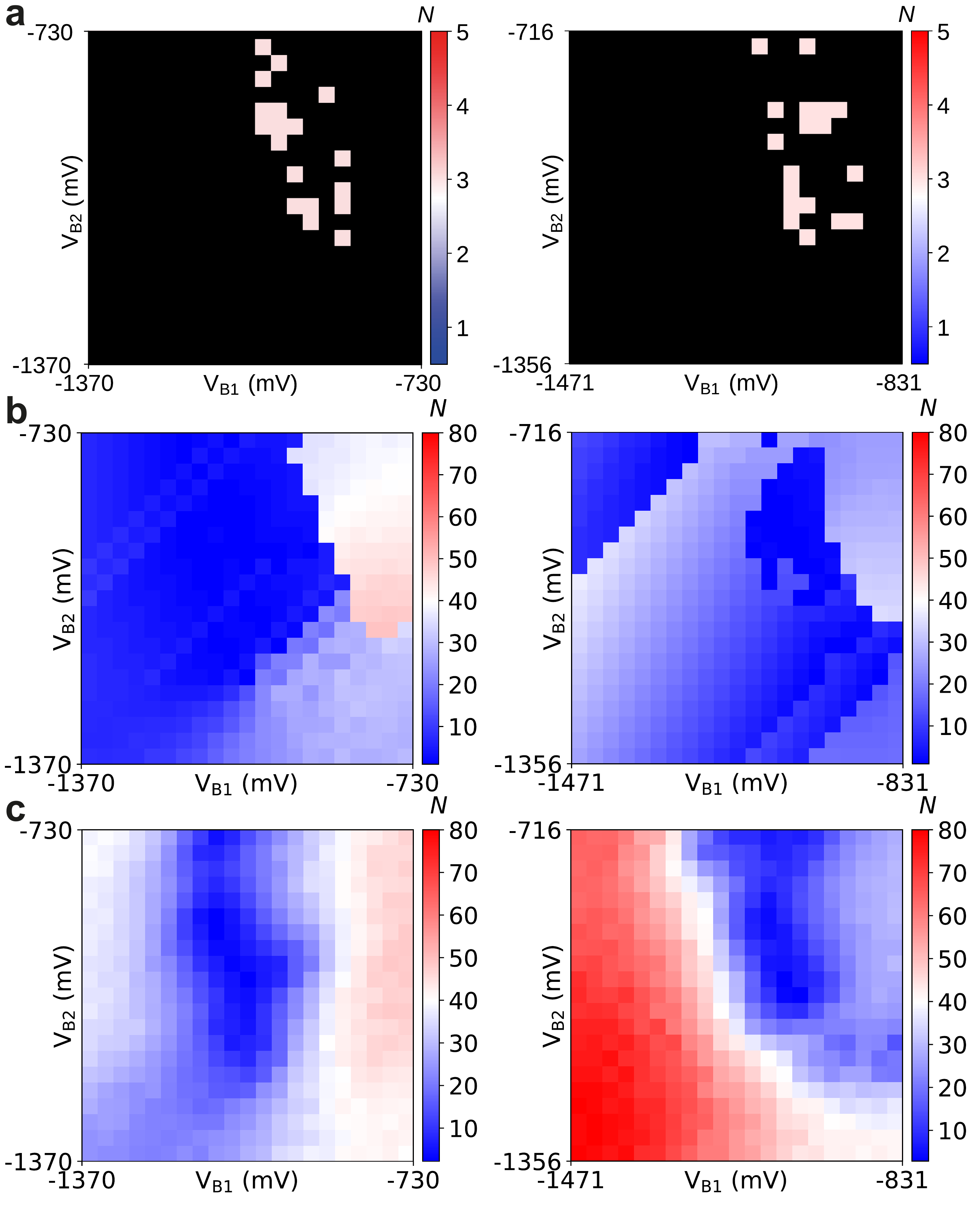}
\par\end{centering}
\centering{}\caption{\textbf{Offline performance distribution.} The performance of different algorithms is evaluated by initiating an algorithm run in each block and estimating $N$ for regions I and II. Black areas indicate that the algorithm failed when initiated at those blocks. \textbf{(a)} Performance distribution (heat-map) for the Nelder-Mead method, \textbf{(b)} the DRL decision agent, and \textbf{(c)} the algorithm with a random decision agent. \label{fig:Performancedistribution}}
\end{figure}

To further illustrate the advantages of our algorithm, we have, for comparison, implemented a Nelder-Mead numerical optimisation method~\cite{Kalantre2019, Zwolak}, an alternative approach not based on reinforcement learning. To ensure a fair comparison with our reinforcement learning method, our implementation of the Nelder-Mead optimisation (see Supplementary Material \ref{sup:nelder-mead} for further details) was terminated when the CNN classified a block as exhibiting bias triangles in the same way as our DRL algorithm, i.e. when the output value of the CNN classifier was greater than $0.5$. In the original implementation, stricter numerical stopping conditions must be met, thereby increasing the number of measurements performed before termination.

The Nelder-Mead, random decision, and DRL decision algorithms were compared offline. We have initiated the algorithms in each block within each gate voltage region and estimated $\bar{N}$, creating a performance distribution or heat map (Fig. \ref{fig:Performancedistribution}). We observe that large areas of gate voltage space which do not exhibit transport features correspond to large flat areas in the optimisation landscape and thus severely limit the Nelder-Mead method. Often the simplex was initiated in these areas and in those cases, the Nelder-Mead algorithm just repeatedly measured the area around the initial simplex. On other occasions, the algorithm moved away from the initial simplex but then became trapped in other areas of the parameter space in which transport features are not present. The method only succeeded in locating bias triangles when it was initiated in the double dot regime. The DRL decision agent's performance is non-uniform as the `pinch-off' regime is less effectively characterised by the agent than the `open' and `single-electron transport' regimes. The performance of the random decision agent is also non-uniform, as it completes the tuning procedure more efficiently when initiated close to the target transport features.

The Nelder-Mead algorithm was also tested online under the same conditions as the DRL and random decision agents. None of $20$ runs succeeded before reaching the predefined maximum number of measurements, $N_\mathrm{max}$, and thus the results are not presented alongside the online results of grid scan, random decision, and DRL decision algorithms in Fig. \ref{fig:PerformanceBenchmark}.

\section*{Discussion}

We have demonstrated efficient measurement of a quantum dot device using reinforcement learning. We are able to locate bias triangles fully automatically from a set of gate voltages defined by a super coarse tuning algorithm \cite{Moon2020}, and in as little as one minute.
Our approach gives a $10$ times speed up in the median time required to locate bias triangles compared with grid scan methods. Our approach is less dependant on the transport regime in which the algorithm is initiated, compared to an algorithm based on a random agent and to a Nelder-Mead numerical optimisation method. We have also demonstrated the statistical advantage of a DRL decision agent over a random decision agent. Our DRL approach is also robust against featureless areas in the parameter space which limit other approaches. While numerical optimisation methods requires time-consuming measurements at each step of the optimisation process, our algorithm uses statistics calculated via pixel sampling to explore the transport landscape. This statistical state representation allows us to efficiently measure the transport regime (or the state of the environment in DRL terms) and avoid over-fitting during agent training. Other options for state representation that go beyond a statistical summary of current values could also be considered. The measurement time remains, however, the dominant contribution in the time required to identify transport features. Fast readout techniques such as radio-frequency reflectometry can be used to reduce measurement times \cite{Crippa2017,Schupp2020,Volk2019,Ares2016,DeJong2019,Jung2012}. 

Our method is inherently flexible and modular such that it could be generalised to automate a variety of efficient measurement tasks. For example, the reward function could be modified so that the agent could learn to locate and score multiple bias triangles within the current map. Furthermore, by retraining the CNN classifier and the DRL agent, the method would be able to locate different types of transport features, such as those observed with charge sensing techniques \cite{Field1993}. Our algorithm could also incorporate other gate electrodes by increasing the action space and retraining. This approach would allow a significantly speed up for super coarse and coarse tuning algorithms. We also expect DRL approaches to scale better than random searches as the dimensionality of the problem increases.

An additional benefit of reinforcement learning is the capacity of the network's policy to be continuously updated. Thereby, the agent's policy can be updated in real-time as the algorithm becomes familiar with a new device. This not only improves the general policy but also means that, over time, the pre-trained agent could learn the particularities of a specific device. To tune large quantum device arrays, due to the increasing dimensionality of the parameter space, DRL could offer a large advantage over conventional heuristic methods. Our quantum dot environment and algorithmic framework offer a valuable resource to develop and test other algorithms and decision agents for quantum device measurement and tuning. Additionally, our dueling deep Q-network methods can be translated to further applications in experimental research.

\subsection*{Acknowledgements}
We acknowledge J. Zimmerman and A. C. Gossard for the growth of the AlGaAs/GaAs heterostructure. This work was supported by the Royal Society, the EPSRC National Quantum Technology Hub in Networked Quantum Information Technology (EP/M013243/1), Quantum Technology Capital (EP/N014995/1), EPSRC Platform Grant (EP/R029229/1), the European Research Council (grant agreement 818751), Graphcore, the Swiss NSF Project 179024, the Swiss Nanoscience Institute, the NCCR QSIT, the NCCR SPIN, and the EU H2020 European Microkelvin Platform EMP grant No. 824109. This publication was also made possible through support from Templeton World Charity Foundation and John Templeton Foundation. The opinions expressed in this publication are those of the authors and do not necessarily reflect the views of the Templeton Foundations. 

\subsection*{Author Contributions}

S.B.O., D.T.L., N.A. and the machine performed the experiments. F.V. contributed to the experiment. V.N. and S.B.O. developed the algorithm in collaboration with M.A.O and D.S. The sample was fabricated by L.C.C., L.Y., and D.M.Z. The project was conceived by V.N. and N.A.. G.A.D.B., V.N., S.B.O and N.A. wrote the manuscript.
All authors commented and discussed the results.

\subsection*{Competing Interests}
The authors declare that they have no competing interests.

\subsection*{Correspondence}
Correspondence and requests for materials
should be addressed to Natalia Ares (email: natalia.ares@materials.ox.ac.uk).




\clearpage

\section*{Supplementary Information}
\subsection{Quantum dot environment}
\label{sup:sup_qde}

In DRL, among the key components is the formal model of the environment with which the agent interacts.
Therefore, we build an environment from the quantum dot device for
training our algorithm and name it the \emph{quantum dot environment} (QDE). The QDE was developed to be compatible with the OpenAI Gym interface \cite{Brockman2016a}. This environment is ready to be used for
benchmarking and training existing DRL algorithms. In addition, this
environment is useful for interested DRL researchers to develop their
methods for improving quantum technologies.

The voltage space of the QDE is delimited by a $\SI{640}{\milli\volt}$ $\times$ $\SI{640}{\milli\volt}$ window defined by the barrier gates. The window is divided in $\SI{32}{\milli\volt}$ $\times$ $\SI{32}{\milli\volt}$ blocks and the agent is initialised in a randomly selected block.

\subsubsection{State} 

A state $s$ is the statistics set, of a block, consisting of the means $\mu$ and standard deviations $\sigma$ of the current in each of the nine sub-blocks. 

Instead of making densely overlapping blocks by a moving kernel horizontally
and vertically, we propose to represent each state by $3$ blocks
per dimension for simplicity. In other words, the image is represented
by $3^{d}$ blocks where $d$ is the dimension. The
dimension $d$ corresponds to the number of gates used. In our setting,
each state in 2 gates includes $9$ blocks as the tensor of $9\times32\times32$
dimensions and 3 gates will have the tensor of $27\times32\times32$.
Examples of the state and blocks using two gates can be found in
Fig. \ref{fig:fig1}.

The design of this state representation will bring two major advantages. This approach ensures that the agent is less prone to over-fitting during training and more robust to experimental noise. The second benefit is for scalability as we can efficiently extend to a higher number of dimensions. While taking a grid scan measurements, which scale exponentially with the number of dimensions, we can use random sampling techniques to obtain convergence in the values of the state representation. This random sampling technique will scale more favourably with higher dimensions. Under this representation, the state includes a statistics vector of $9\times2$ dimensions.

\subsubsection{Action} 

Our action space includes increasing ($+$) or
decreasing ($-$) each gate voltage. We have specially designed two actions  to modify both gates simultaneously. In higher dimensional setting, such as controlling $d>2$ gates, this action space can be generalised wherein the number of actions is $2\times d+2$.

\subsubsection{Reward} 

The reward function is carefully constructed to force the agent to learn to navigate through the voltage landscape and identify bias triangles using the fewest $N$ measurements. We follow the popular Taxi-v2 environment\footnote{\href{https://gym.openai.com/envs/Taxi-v2/}{https://gym.openai.com/envs/Taxi-v2/}} to design the reward scores. We summarise the components of the reward function in Table \ref{tab:reward}. We note that the utility of the agent is robust with respect to different magnitudes of these scores, provided that the detection of a pair of bias triangles receives a much higher score than block measurement steps.

\begin{table}[h!]
    \centering
       \caption{Summary of the reward function.}
    \begin{tabular}{|l|l|l|}
         \hline
         \textbf{Instance} & \textbf{Reward} & \textbf{Termination} \\
         \hline
         \hline
         Each block measured & -1 & False
         \\
         \hline
         Bias triangle detected & +10 & True \\
         \hline
         $N$ equal to $N_\mathrm{max}$ & -10 & True \\
         \hline
         \end{tabular}
 
    \label{tab:reward}
\end{table}

We assign high reward to our target state of bias-triangles. We encourage the algorithm to find the bias-triangles
using the fewest number of measurement by designing the reward score
as follows. We assign the highest reward $r=+10$ to the bias-triangles
location. Then, other
states will take $r=-1$. The maximum number of steps per episode during training
is set as $300$. Beyond this threshold, if the algorithm cannot find the bias-triangles, it will terminate and assign $r=-10$. The
maximum number of steps controls how far away from a starting point the device-measurement can go.

\begin{figure}[!h]
\begin{centering}
\includegraphics[width=0.99\columnwidth]{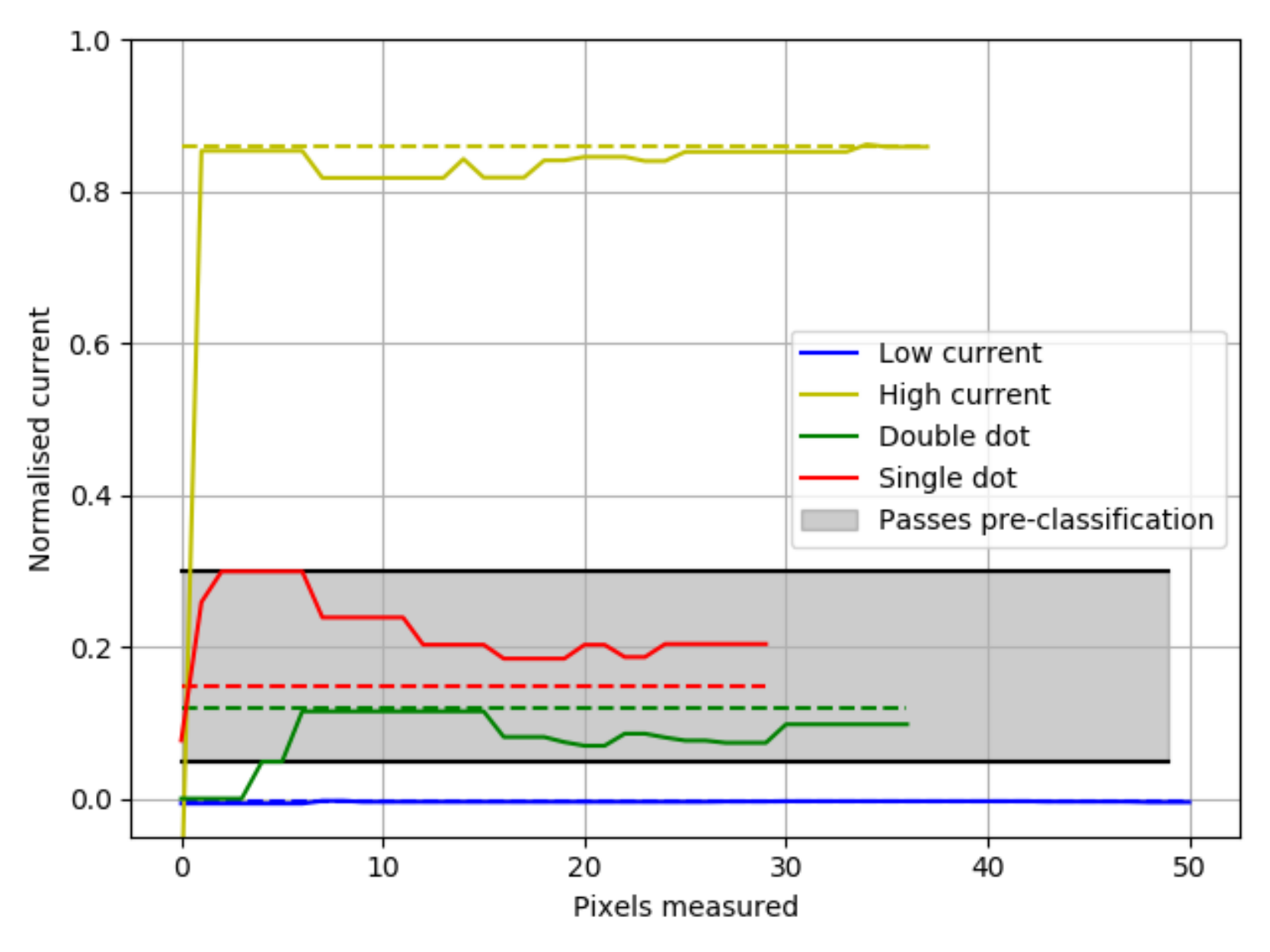} 
\par\end{centering}
\centering{}\caption{\textbf{Pixel sampling convergence} for the random pixel measurement in sub-blocks taken from pinch-off (low current), open (high current), single dot, and double dot regimes. The threshold values set for the pre-classification tools are indicated. Therefore, any sub-block within the grey shaded region will pass the pre-classification. The dashed lines represent the true means after measuring the sub-blocks using a $\SI{1}{\milli\volt} \times \SI{1}{\milli\volt}$ resolution scan. \label{fig:PixelSampling}}
\end{figure}

\subsection{Random Pixel Measurement and Pre-classifier}
\label{sup:sup_random_pixel}

To assess the state of a block, the algorithm first conducts a random pixel measurement. Pixels are repeatedly sampled at random from the block and the statistics are calculated for each sub-block until convergence. The convergence of both the mean and standard deviation of each sub-block must be satisfied before the measurement is terminated. The convergence is accepted if the mean change in the values of the state representation is less than a threshold percentage (one percent threshold for this paper) of the state representation prior to the update. The state vector is then assessed by the pre-classification stage. If the mean current values, of any of the sub-blocks, falls between the threshold values, calculated using the initialisation one-dimensional scans, then the block is pre-classified as a boundary region. The convergence of the normalised mean current of a sub-block, in the random pixel sampling measurement method is shown in Fig. \ref{fig:PixelSampling}, for low and high current, as well as single and double dot sub-blocks. The sub-blocks with single and double dot transport features fall within the pre-classifier threshold values and therefore, in the full algorithm, would be measured using a grid scan and evaluated by the CNN binary classifier.

Satisfactory convergence for a block is achieved in fewer than $50$ pixel measurements in all regimes, compared to the $1,024$ pixels measured in a grid scan of the block. This represents a huge improvement in measurement efficiency and the evaluation of a state of the DRL agent.

\subsection{CNN Binary Classifier}
\label{sup:sup_cnn}
A Convolutional Neural Network (CNN) \cite{Krizhevsky2012,Lecun2015}  is a multilayered neural network with a special architecture to detect complex patterns. To decide whether the agent has found bias triangles in a given block, the algorithm uses a CNN as a binary classification tool. If the CNN outputs a value greater than a $0.5$ threshold, corresponding to the classification of a pair of bias triangles, then the algorithm is terminated. An optimisation of the threshold value could enable us to reduce the number of false positive classifications.

\subsubsection{Network Architecture}

\begin{figure}
\includegraphics[width=0.99\columnwidth]{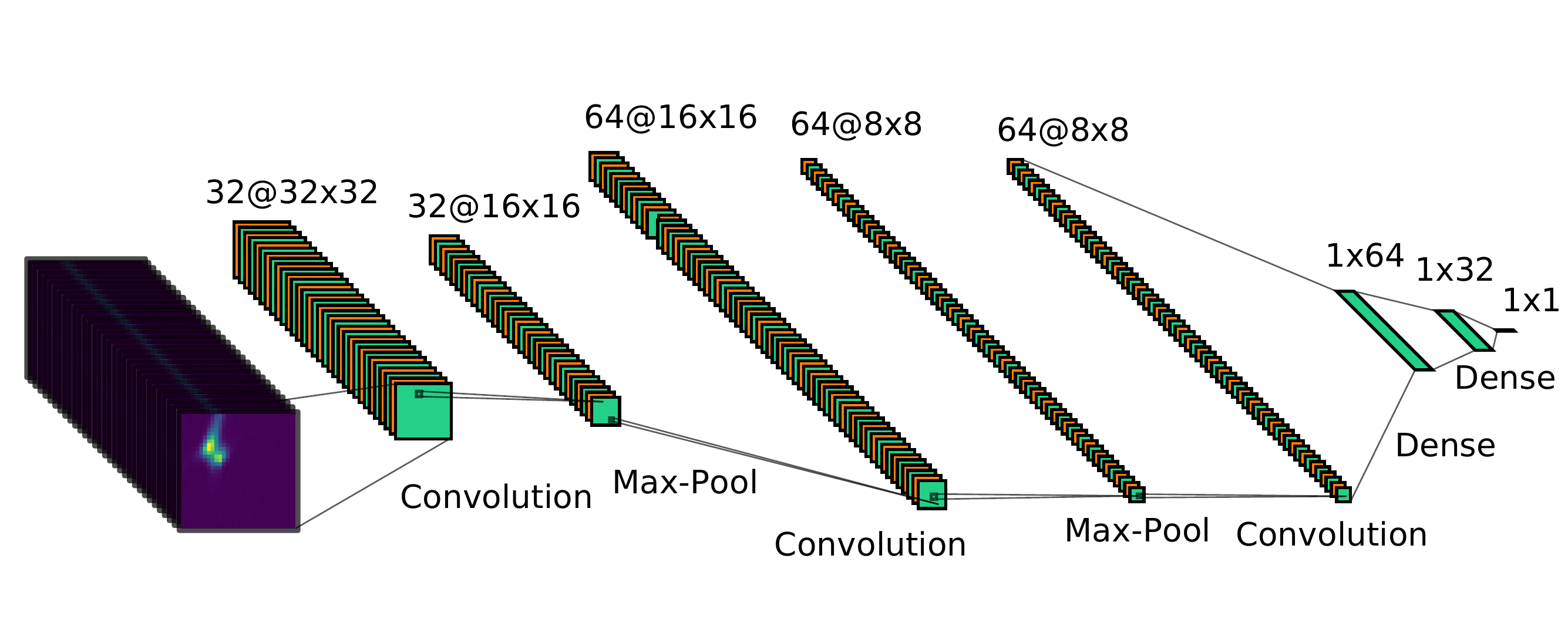}
\caption{\textbf{CNN binary classifier architecture.} \label{fig:cnn_arch}}
\end{figure}

We summarise in Fig. \ref{fig:cnn_arch} the network architecture and hyperparameters used. There are a total of $320065$ trainable parameters. The convolutional layers have a Rectified Linear Unit (RELU) activation function, while the dense layers have and Exponential Linear Unit (ELU) activation function. The final, output, layer has a Sigmoid activation function.

\begin{table}
\caption{CNN binary classifier confusion matrix \label{tab:confusion}}

\begin{tabular}{|l|l|}
\hline 
\multicolumn{2}{|c|}{Confusion Parameters}\tabularnewline
\hline 
True positive  & $18 \%$\tabularnewline
\midrule
False positive &  $4 \%$\tabularnewline
\midrule
True negative  & $76 \%$\tabularnewline
\midrule
False negative &  $2 \%$\tabularnewline
\hline
\hline 
F-measure & $ 85 \%$\tabularnewline
\hline 
\hline 
Accuracy &  $ 94 \%$\tabularnewline
\hline 
\end{tabular}
\end{table}

\subsubsection{Confusion Matrix}

The CNN was trained over $10$ epochs using $11425$ data points in the training set, $4896$ in the validation set and $6994$ in the test set. Data was augmented by applying rotations. We trained the network using an Adam optimiser \cite{Kingma2015} and a binary cross-entropy loss function. Regularisation was achieved using a L2 regulariser, set to $0.0001$, as well as drop-out for the dense layers, set at $0.1$. In Table \ref{tab:confusion} we present a summary of the prediction results on the test set of the binary classification problem. The confusion parameter representation is useful for analysing the types of error that a classifier typically makes. The F-measure and accuracy are other commonly used metrics to analyse the efficacy of a binary classification tool.

\subsubsection{Positive Examples}

\begin{figure}
\begin{centering}
\includegraphics[width=0.9\columnwidth]{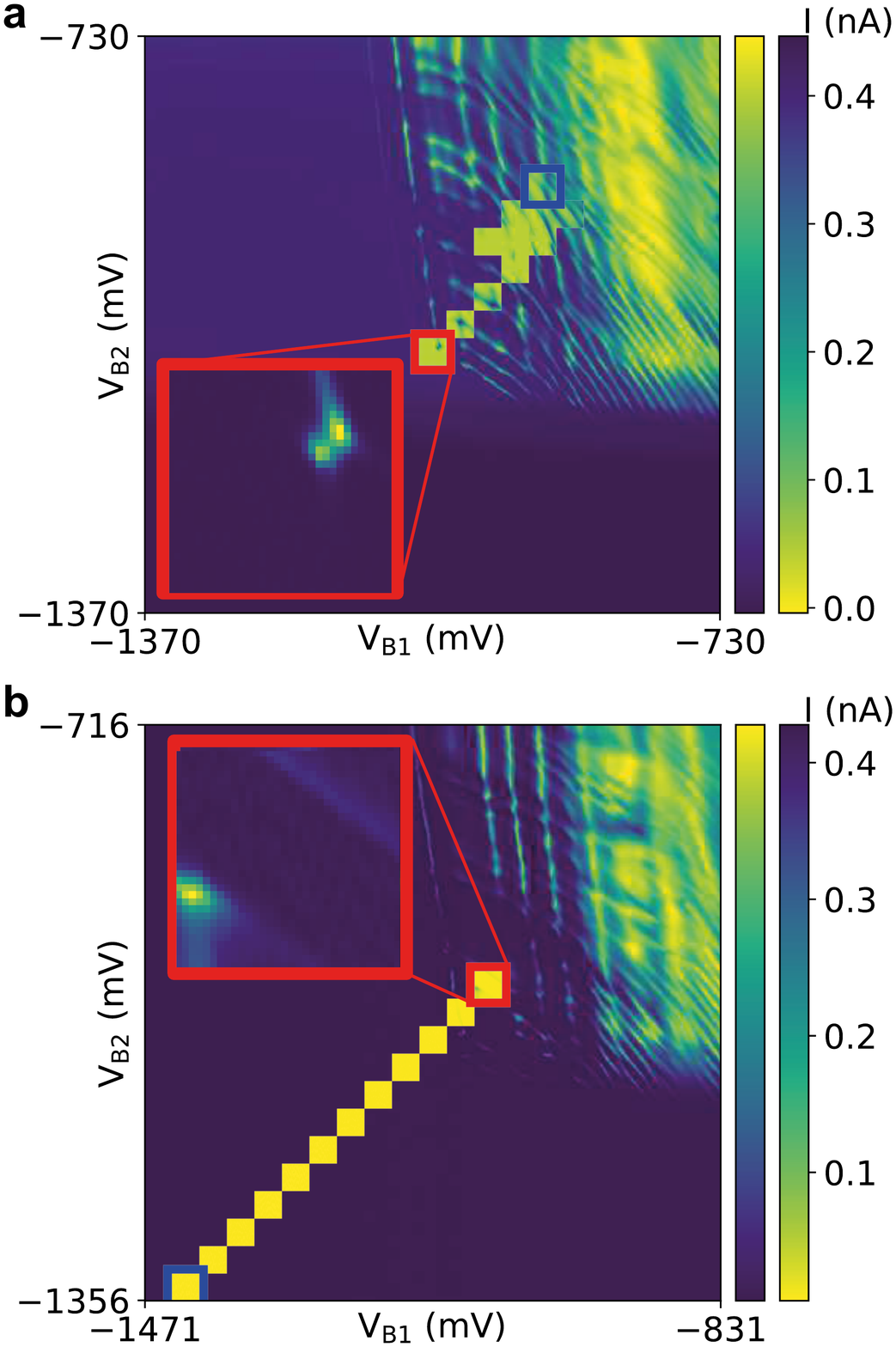} 
\par\end{centering}
\centering{}\caption{\textbf{Example trajectories}. (\textbf{a,b}) Example trajectories from Fig. \ref{fig:PerformanceBenchmark} \textbf{a, b}, respectively, with the insets showing the bias triangles which triggered the stopping condition for the algorithm. \label{fig:trajectory}}
\end{figure}

As the DRL agent navigates through the environment, the algorithm evaluates each block using the pre-classification protocol. If the block passes the pre-classifier stage, a grid scan of the block is measured and the CNN binary classification tool is used to evaluate the block. If the CNN positively classifies the block as containing bias triangles the algorithm is terminated and the run treated as successful. In Fig. \ref{fig:trajectory} we show the blocks that were positively classified by the CNN, causing the algorithm to terminate during the real-time measurements.

\subsection{DRL Decision Agent}
\label{sup:sup_drl_agent}

\begin{table}
\caption{Deep Reinforcement Learning Architecture \label{tab:Deep-Reinforcement-Learning_architecture}}

\begin{tabular}{|l|l|}
\hline 
\multicolumn{2}{|c|}{\textbf{Hyper-parameters Used}}\tabularnewline
\hline 
Discount factor & $0.5$\tabularnewline
\midrule
Optimizer & Adam\tabularnewline
\midrule
Number of episodes & $10000$\tabularnewline
\midrule
Mini batch-size & $32$\tabularnewline
\midrule
Decay rate in $\epsilon$ greedy & $1e^{-4}$\tabularnewline
\midrule
Replay buffer & $20000$\tabularnewline
\midrule
PER-$\beta$ (start, final, no steps) & $(1.0,0.6,1000)$\tabularnewline
\midrule
Learning rate & $2.5e^{-6}$\tabularnewline
\midrule
FC Layers & $128,64,32$\tabularnewline
\midrule
FC Layers (Dueling) & $64,1$\tabularnewline
\hline 
\end{tabular}
\vspace{20pt}
\end{table}

\subsubsection{Network Architecture}

\begin{figure}
\includegraphics[width=1\columnwidth]{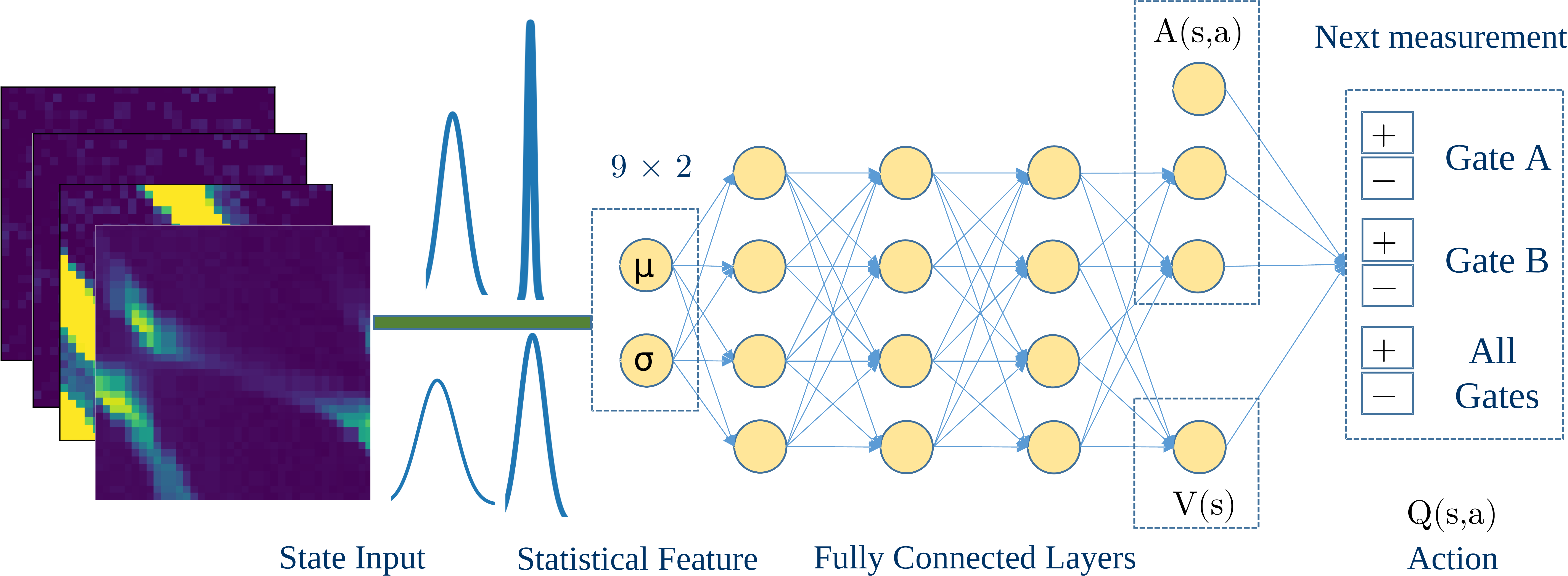}
\caption{\textbf{Summary of DRL framework.} Our deep reinforcement learning framework using a statistical state representation. \label{fig:DRL-Framework-FC-3D}}
\end{figure}

We first summarise the
network architecture and hyperparameters used in Table \ref{tab:Deep-Reinforcement-Learning_architecture}.
We further illustrate the model architecture in Fig. \ref{fig:DRL-Framework-FC-3D}.

\subsubsection{Training}

\begin{figure}
\includegraphics[width=1\columnwidth]{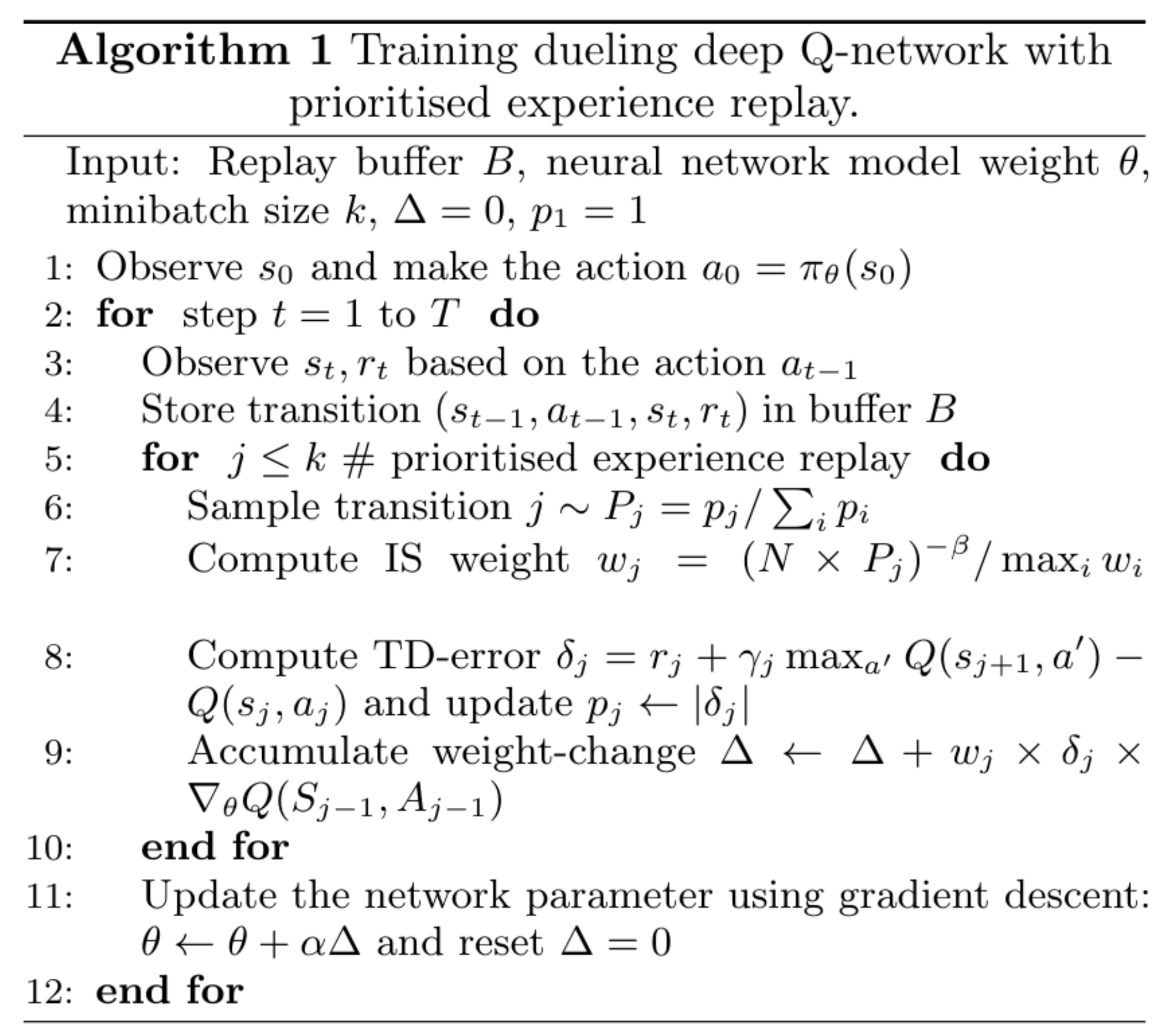}
\caption{\textbf{Pseudocode.} Training the dueling deep Q-network with prioritised experience replay. \label{fig:algorithm_pseudocode}}
\end{figure}

We present the pseudo code for the training of the DRL decision agent with prioritised experience replay in Fig. \ref{fig:algorithm_pseudocode}. The training process starts as follows. An agent initially will make random action choices to gain experiences which will be stored in a replay buffer $B$. From this buffer, the data sample will be randomly selected at a rate proportional to the temporal difference (TD) error. Particularly, it prefers to pick samples with unexpected transitions since these contain more information to learn than from others samples. Then, the neural network will be updated given such `unexpected' samples to improve the networks policy for the next iterations.

\begin{figure}
\includegraphics[width=0.92\columnwidth]{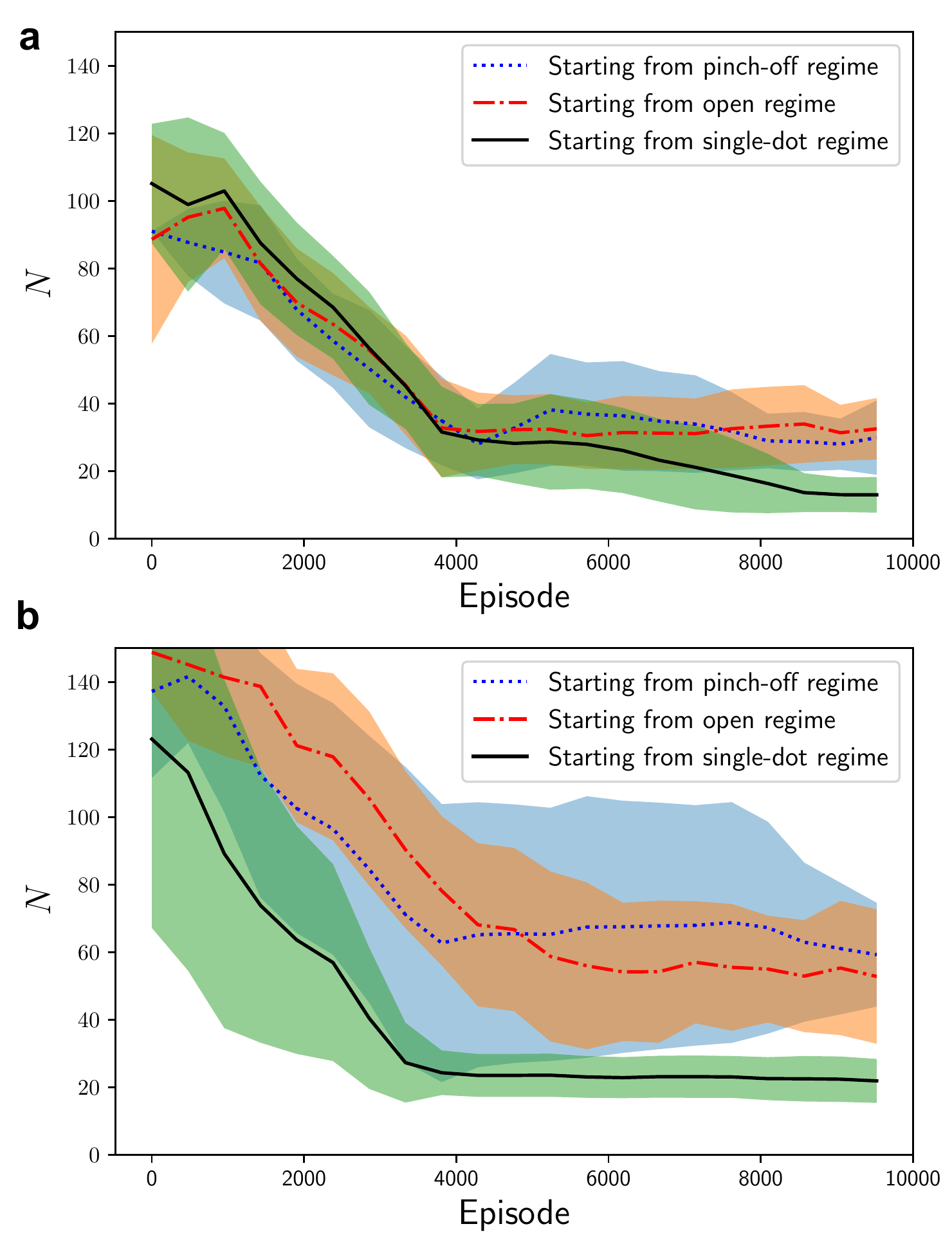}
\caption{\textbf{Training convergence.} We evaluate the performance of our DRL agent during a learning process. The region of gate voltage space is different to the ones explored in the main text. The bias triangles are labelled in advance by experts. Lines show the mean $N$ with uncertainty bounds for different starting locations. 
The performance test in \textbf{a} was run on the same device in which training was performed, while in \textbf{b}, the test was run on a different device to the one used for training. \label{fig:NoMeasurement_DifferentLocs}}
\end{figure}

We then illustrate the learning process of our DRL agent by showing the $N$ measurements required to locate bias triangles as a function of the number of training episodes (Fig. \ref{fig:NoMeasurement_DifferentLocs}). This test, which was run in different regions of gate voltage space to the ones explored in the main text, was performed to assess the learning rate of the DRL network. Bias triangles were labelled in advance by human experts and the CNN and pre-classifier modules were not used. Unsurprisingly, when we test the performance of the agent in the same device in which it is trained, less number of learning episodes are required compared to when we perform the test run in a different device. However, in order to be robust against device variability, training and testing have to be run in different devices. 

\subsubsection{Performance}

We summarise the online performance, in Table \ref{tab:online_performance summary} and the offline performance in Table \ref{tab:offline_performance summary}, of the DRL decision agent with respect to the random decision agent. We use the two-tailed Wilcoxon signed rank test \cite{Wilcoxon1945} to assess the null hypothesis that the DRL and Random agent's performances are drawn from the same distribution. The p-value, given in Table \ref{tab:online_wilcoxon} and Table \ref{tab:offline_wilcoxon} for online and offline tests respectively, represents the confidence in the null hypothesis. The null hypothesis can only be rejected with confidence, at a level of $2\%$, in the online results in the case of $\bar{N}$ in region II. For the offline results, the null hypothesis can be rejected for $\bar{N}$ in both regions. A one-tailed Wilcoxon signed rank test demonstrates that the median of the differences $(\bar{N}_{\mathrm{DRL}}-\bar{N}_{\mathrm{Random}})$ is less than zero. We can therefore conclude that the DRL agent offers a statistically significant advantage over the random agent.

\begin{table}
    \centering
     \caption{Summary of the \textbf{online} performance of the DRL and Random decision agents online in the two parameter regimes. The performance metrics used are the number of blocks measured, $N$, before identifying a pair of bias triangles and the corresponding lab time.}
    \begin{tabular}{|l|l|l|}
         \hline
         Agent & \textbf{DRL} & \textbf{Random} \\
         \hline
         \hline
         \textbf{Region I median lab time (s)} & 932 &  683\\
         \hline
         $10\%$ percentile time (s) &  228 & 222  \\
         $90\%$ percentile time (s) & 2430 & 4844 \\
         \hline
         \hline
         \textbf{Region II median lab time (s)} & 822 & 989 \\
         \hline
         $10\%$ percentile time (s) & 181 & 349  \\
         $90\%$ percentile time (s) & 1766 & 1500 \\
         \hline
         \hline
         \textbf{Region I \scriptsize{$\bar{N}$}} & 41 & 54
         \\
         \hline
         $10\%$ percentile $N$ & 9 & 15  \\
         $90\%$ percentile $N$ & 104 & 135 \\
         \hline
         \hline
         \textbf{Region II \scriptsize{$\bar{N}$}} & 32 &  85 \\
         \hline
         $10\%$ percentile $N$ & 10 & 50   \\
         $90\%$ percentile $N$ & 94 &
         143 \\
         \hline
         \end{tabular}
   
    \label{tab:online_performance summary}
\end{table}

\begin{table}
    \centering
     \caption{Summary of the Wilcoxon signed rank test analysis on the \textbf{online} performance of the DRL and Random decision agents in the two parameter regions. The performance metrics used are the number of blocks measured, $N$, before identifying a pair of bias triangle and the corresponding lab time.}
    \begin{tabular}{|l|l|}
         \hline
         & \textbf{Wilcoxon signed rank p-value}\\
         \hline
         \hline
         \textbf{Region I lab time} & 0.72 \\
         \hline
         \textbf{Region II lab time} & 0.58 \\
         \hline
         \textbf{Region I $N$} & 0.72 \\
         \hline
         \textbf{Region II $N$} &  0.02 \\
         \hline
         \end{tabular}
   
    \label{tab:online_wilcoxon}
\end{table}

\begin{table}
    \centering
       \caption{Summary of the \textbf{offline} performance of the DRL and Random decision agents online in the two parameter regions. The performance metrics used are the number of blocks explored, $N$, before identifying a pair of bias triangles (the lower the $\bar{N}$, the better the performance). For offline experiments, lab times are not a performance metric to be considered.}
    \begin{tabular}{|l|l|l|}
         \hline
         & \textbf{DRL agent} & \textbf{Random agent} \\
         \hline
         \hline
         \textbf{Region I \footnotesize{$\bar{N}$}} & 6 & 22
         \\
         \hline
         $10\%$ percentile $N$ & 1 & 2  \\
         $90\%$ percentile $N$ & 64 & 46 \\
         \hline
         \hline
         \textbf{Region II \footnotesize{$\bar{N}$}} & 17 & 30  \\
         \hline
         $10\%$ percentile $N$ & 2 & 3  \\
         $90\%$ percentile $N$ & 31 & 101 \\
         \hline
         \end{tabular}
 
    \label{tab:offline_performance summary}
\end{table}

\begin{table}
    \centering
     \caption{Summary of the Wilcoxon signed rank test analysis on the \textbf{offline} performance of the DRL and Random decision agents in the two parameter regions. The performance metrics used are the number of blocks measured, $N$. For offline experiments, lab times are not a performance metric to be considered.}
    \begin{tabular}{|l|l|}
         \hline
         & \textbf{Wilcoxon signed rank p-value}\\
         \hline
         \hline
         \textbf{Region I $N$} & $<$0.001 \\
         \hline
         \textbf{Region II $N$} &  $<$0.001 \\
         \hline
         \end{tabular}
   
    \label{tab:offline_wilcoxon}
\end{table}

\subsubsection{Policy}

\begin{figure*}

\includegraphics[width=1.25\columnwidth]{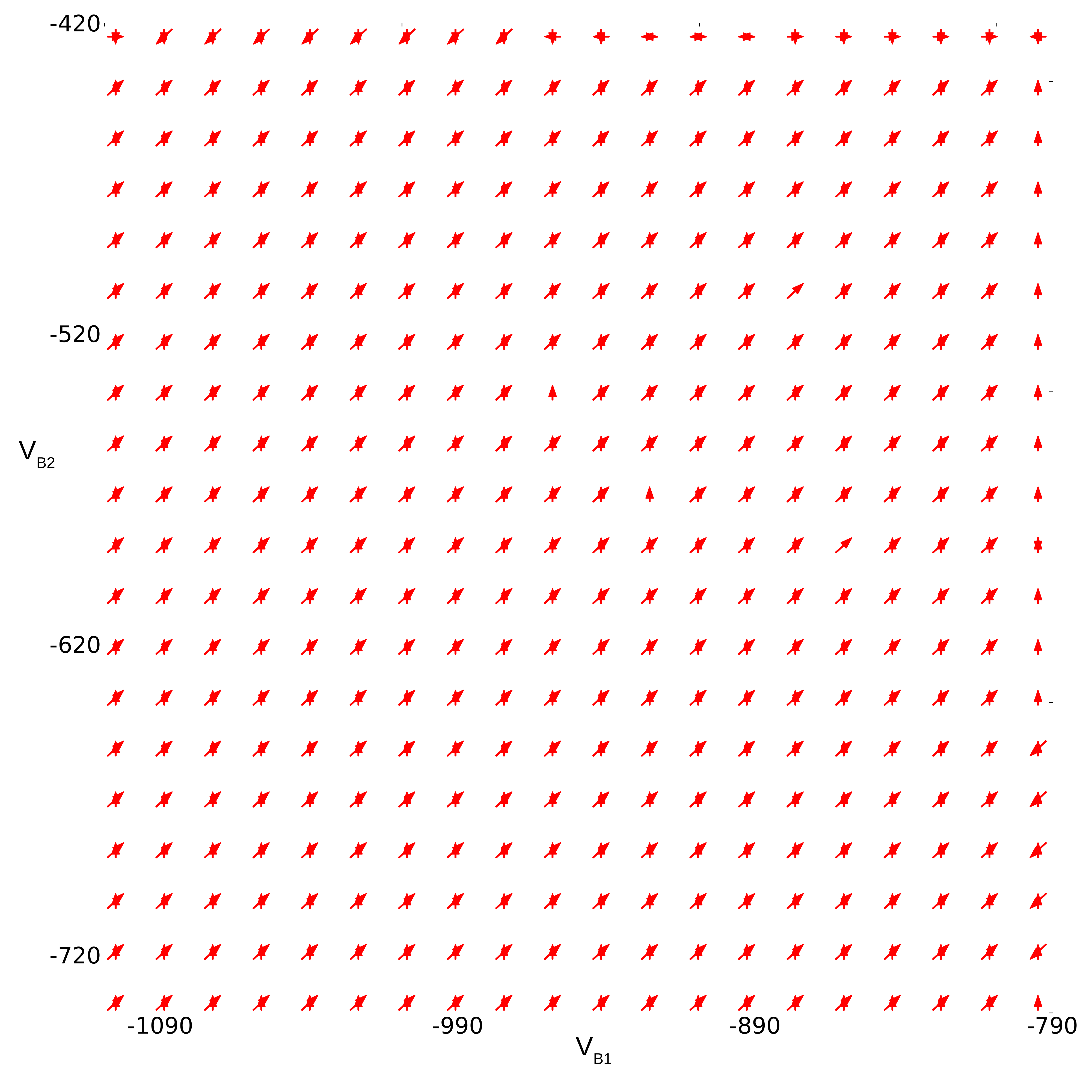} \\
\includegraphics[width=1.25\columnwidth]{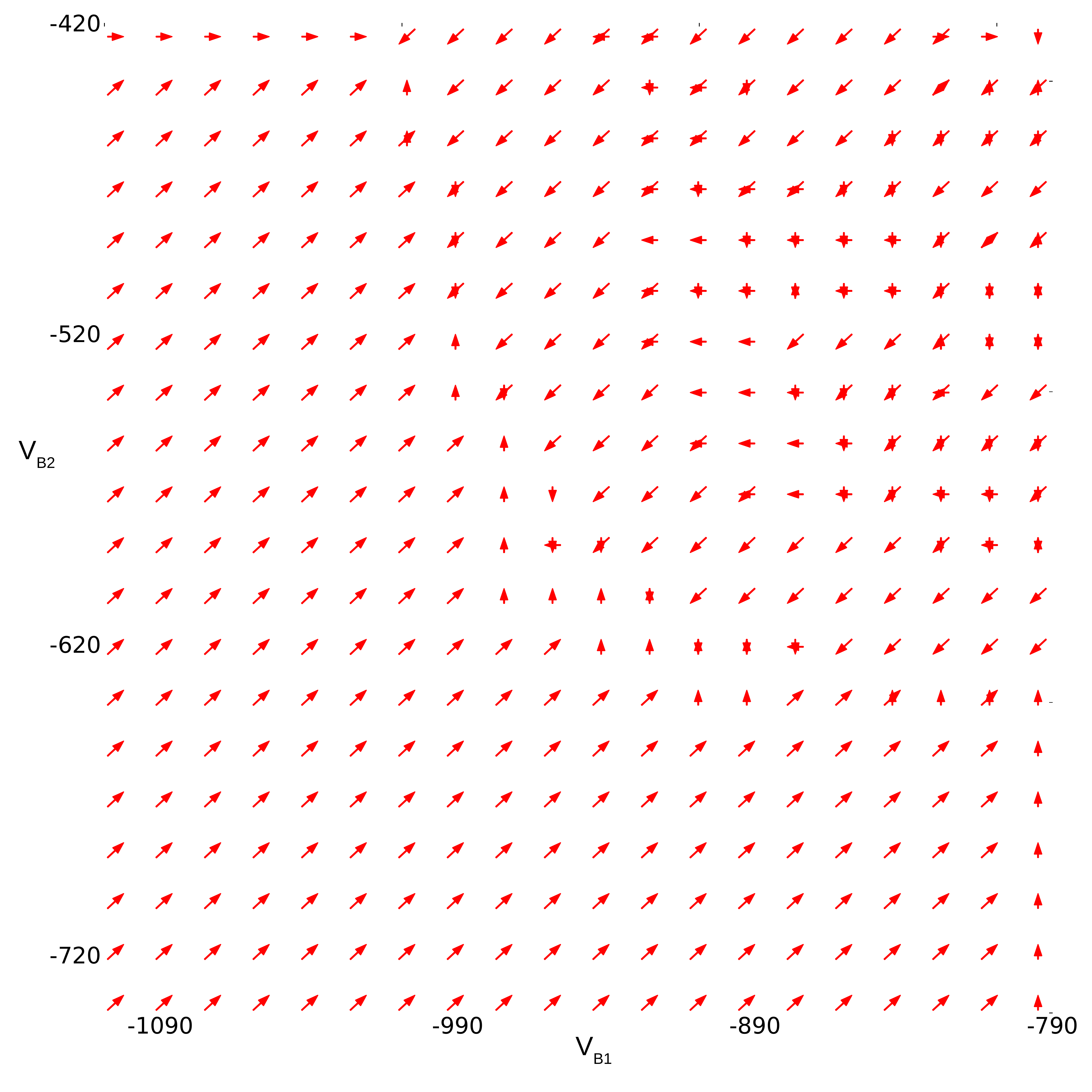}
\caption{\textbf{Optimal policies.} We plot the optimal policies learned at \textbf{Top} early stage and \textbf{Bottom} later stage of the training process. The arrows indicates, at the given location, the optimal direction to move. Two arrows represent two probable actions which both with high chances of being optimal. At the early stage of the training, the agent's policy has more uncertainty regarding which action to select. While at the later stage, after being trained, it consistently makes the optimal decision.} \label{fig:OptimalPolicy}
\end{figure*}

In the reinforcement learning context, a policy defines what an agent does to accomplish a task. We present the optimal policies at different training stages in Fig. \ref{fig:OptimalPolicy} wherein we use arrows to indicate the action, i.e., the direction to move in the gate voltage space to perform the next measurement. The algorithm learns that it should move towards more positive gate voltages if the state is pinch-off (low-current) or go towards more negative gate voltages if the state is the open regime (high-current). 

\subsection{The Nelder-Mead numerical optimisation method}
\label{sup:nelder-mead}

We construct a fitness function by taking the \(L^{2}\)-norm of a probability vector defining a difference metric between the current state, i.e. a given transport regime, and the target state or transport regime. In slight variance from the implementation in \cite{Zwolak}, we have defined the probability vector of the current state as \(\left(p(s), 1 - p(s)\right)^{T}\) and the target vector defined as \((1, 0)^{T}\). $s$ is a coordinate in gate voltage space and this coordinate's fitness value is calculated, as above, by evaluating the CNN prediction $p(s)$ of the probability that a window ($\SI{32}{\milli\volt}$ \(\times\) $\SI{32}{\milli\volt}$) defined around $s$ contains bias triangles. Thus, in single and double dot transport regimes, the value of $p(s)$ should be higher than in the pinch-off and open regimes. The value of the \(L^{2}\)-norm should have minima at the locations of bias triangles. The Nelder-Mead numerical optimisation method, with two gate voltages as free parameters, then automated the location of these minima. This method converges on local minima, in \(n\) dimensions, by evaluating a set of \(n+1\) test coordinates within the optimisation landscape, called a simplex. We defined the initial simplex similarly to \cite{Zwolak}, as the fitness value of the starting (\(s\)) and two additional coordinates obtained by reducing the voltage on each of the barrier gate voltages one at a time by $\SI{75}{\milli\volt}$.

\end{document}